\documentclass[10pt, conference, compsocconf]{IEEEtran}
\ifCLASSINFOpdf
  \usepackage[pdftex]{graphicx}
\else
\fi
\usepackage[cmex10]{amsmath}

\usepackage{bibspacing}
\usepackage{multirow}
\usepackage{array}
\usepackage{caption,subcaption}
\begin{document}
%
\title{A Survey on Simulation Tools and Testbeds for Cognitive Radio Networks Study}


\author{\IEEEauthorblockN{Qi Dong${^\dagger}$, Yu Chen${^\dagger}$, Xiaohua Li${^\dagger}$, Kai Zeng${^\ddagger}$}
\IEEEauthorblockA{${^\dagger}$Dept. of Electrical and Computing Engineering, Binghamton University, SUNY, Binghamton, NY 13902, USA\\
${^\ddagger}$Dept. of Electrical and Computing Engineering, George Mason University, Fairfax, VA 22030, USA}
 E-mail: \{qdong3, ychen, xli\}@binghamton.edu, kzeng2@gmu.edu
}

\maketitle

\begin{abstract}
Efficient utility of radio spectrum has been a hot topic as the wireless communication spectrum is a precious resource. The past decade has witnessed intensive research in spectrum sharing techniques. Most of the techniques are based on cognitive radio networks (CRNs) because cognitive capabilities are essential for optimizing spectrum efficiency and guaranteeing safe coexistence in the presence of the spectrum uncertainty. However, due to the high complexity of the problem, most research has been limited to theoretical analysis. It is non-trivial to build a simulator that is capable of carrying out a comprehensive experimental study. In this paper, a survey is conducted to provide a big picture of the available simulators in CRNs research. By illustrating their major functionalities, the insight enables researchers to select tools that match their needs. In addition, with a better understanding of the advantages and constraints, this survey aims at providing a guideline for simulator designers who have been trying to meet the requirements from the CRN research community.
\end{abstract}

\begin{IEEEkeywords}
Cognitive Radio Networks (CRNs), Spectrum Sharing, Wireless Communication, Simulators, Testbeds.
\end{IEEEkeywords}

\IEEEpeerreviewmaketitle

\section{Introduction}
\begin{figure}[t]
	\centering
		\includegraphics[width=0.45\textwidth]{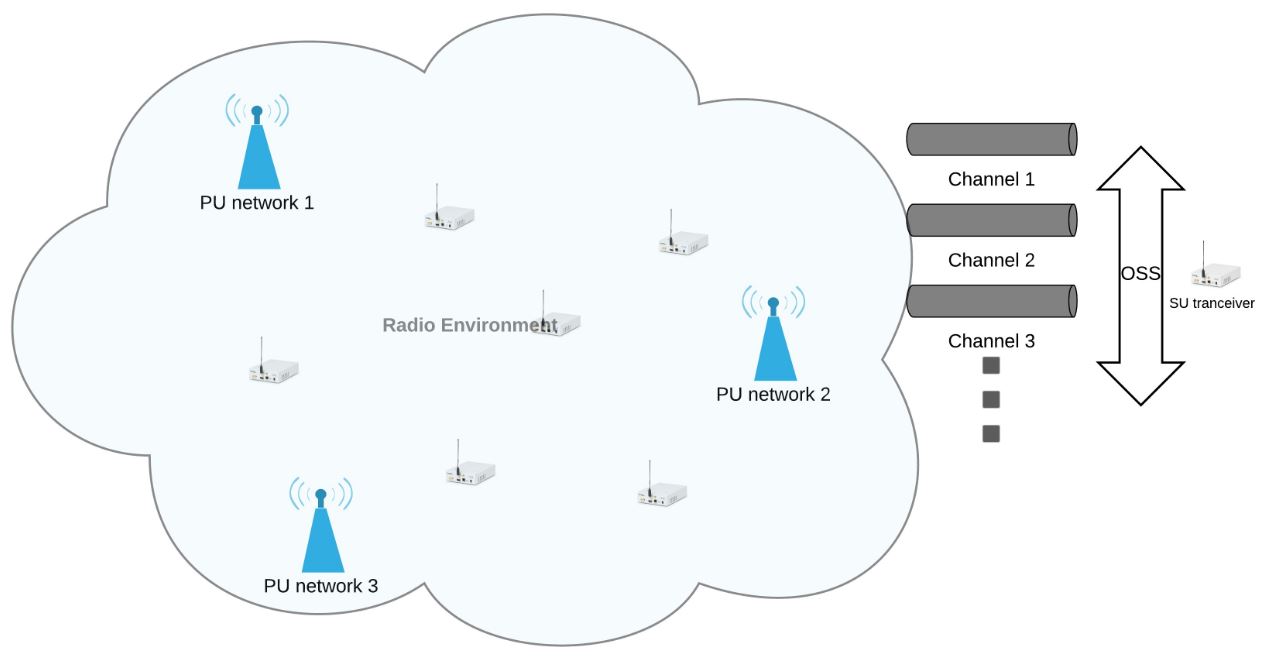}
	\caption{CRN on PU networks.}
	\label{fig:outline}
	\vspace{-18pt}
\end{figure}

Recent years, the increasing demands for wireless communication spectrum have inspired a lot of research toward new technologies for dynamic use of frequency bands. Cognitive radio, which is built upon software-defined radio (SDR) \cite{Haykin2005}, has become a widely recognized approach to exploit spectrum bands. Due to the shortage of spectrum resource, Cognitive radio network (CRN) is a wireless network configured to coexistence with legitimate wireless communication parties, which are tagged as primary users (PUs); cognitive radios (CRs), also known as secondary users (SUs), share the same spectrum resource with PUs without introducing major interference. In CRNs, SUs can automatically sense the licensed channels authorized to PUs, and intelligently utilize the spectrum resource based on smart policies. Essentially, CR technology is a combination of traditional wireless network application and novel frequency resource management. Thus, on one hand, CRs are composed by wireless network protocols layer by layer to construct a robust communicational network; on the other hand, CRs are extending frequency resource usage by adopting opportunistic spectrum sharing (OSS) manner, as shown by Fig. \ref{fig:outline}.

A brief work process of CRNs includes spectrum sensing, spectrum analysis and spectrum decision. Since CRNs aim at exploiting spectrum resource, SUs need to be aware of the occupation status of spectrum by sensing the radio environment. To achieve that, spectrum hole (the spectrum band that can be temporarily used) detection technique helps to find proper communication bands. It requires SUs being able to filter the desired spectrum band and measure its energy level. When the spared spectrum bands are discovered, which shows different characteristics according to the time varying radio environment, a comprehensive analysis is needed considering spectrum interference, path loss, wireless link errors, link layer delay and holding time \cite{Akyildiz:2006aa}. According to synthetic spectrum information and user requirements, the spectrum resource distribution rule should be drafted with consideration of unstable channel status, spectrum hand-off scheme and quality of service (QoS) guarantee.

In addition, due to the complexity and open feature of the system, CRNs are vulnerable to several types of malicious attacks, including not only the traditional network threat such as jamming, eavesdropping, and MAC layer attack, but also certain CRN specific attacks such as control channel jamming, sybil attack, black hole attack and some newly developed types of attacks \cite{Esch:2012aa}, for example, the Primary User Emulation (PUE) Attacks \cite{dong2017exploration}, \cite{dong2018explore}, \cite{dong2017anomaly}.

To obtain deeper insight of CRNs and to design more robust protocols, it is essential to have a thorough understanding of CRNs via intensive experimental analysis. Generally, there are two types of CRN experimental analysis methods: Computer assisted simulation (can model hypothetical objects or network activities on a computer) \cite{Pan:2008aa}, and platform based emulation (can model real radio frequency environment and networks via dedicated designed hardwares). Compared to platform based experimental analysis, computer assisted simulation gains wider utilization in CRN experimental analysis due to several merits: 

\begin{enumerate}
   \item computer assisted simulators are more handy and cheaper; CRN platforms usually cost hundreds even thousands for a single board, while many simulators are open source with free of charge or have academic versions can be downloaded from the Internet; 
   \item Most of the computer assisted simulators are attached with better technique supports; many network simulators have been developed for years and built up with rich wireless communication libraries; although some commercial CRN platforms are technically well supported, they still suffer from shortage of module libraries due to the length of development; and
   \item The computer assisted simulators are more suitable for large scaled CRN experiment with less limitations; it is more applicable to adjust wide range of CRN parameters in simulators, instead of modifying platform hardware design; some radio scenarios can be hardly modeled by radio platforms in real-life, while computer assisted simulation is totally applicable.
\end{enumerate}

However, choosing a user-friendly and powerful simulator is still a non-trivial issue to a lot of researchers because: 

\begin{enumerate}
   \item simulators are maintained by either commercial companies or open source communities; historical technical development and current technique support (CRN patches, libraries, etc.) situations vary greatly among different simulators; and 
   \item CRN is a sophisticated system and certain researches usually undesired to cover complete functionalities; some simulators may expertize to certain CRN activity simulations, others may be used for full CRN design.
\end{enumerate}

This survey aims to provide an overview of current available CRN simulators. By illustrating their major functionalities and substantiate their usage by providing rudimentary case examples, the insight will help researchers to choose proper tools that match their needs. The better understanding of the advantages and constraints will also provide a guideline for simulator designers to meet the requirements from the CRN research community.

The remainder of this paper is structured as the follows. Section \ref{sec:requirements} describes fundamental characters of CRN activities and corresponding simulators; a general CRN simulator classification is also discussed. Section \ref{sec:radio} presents important radio generators in CRN study. Non-agent-based CRN simulators and agent-based CRN simulators are discussed in section \ref{sec:non} and section \ref{sec:agent} respectively. Section \ref{sec:conclusion} casts a short summary of the paper.

\section{Features of CRN simulators}
\label{sec:requirements}

\subsection{CR architecture and modeling basics}
\begin{figure}[t]
	\centering
		\includegraphics[width=0.48\textwidth]{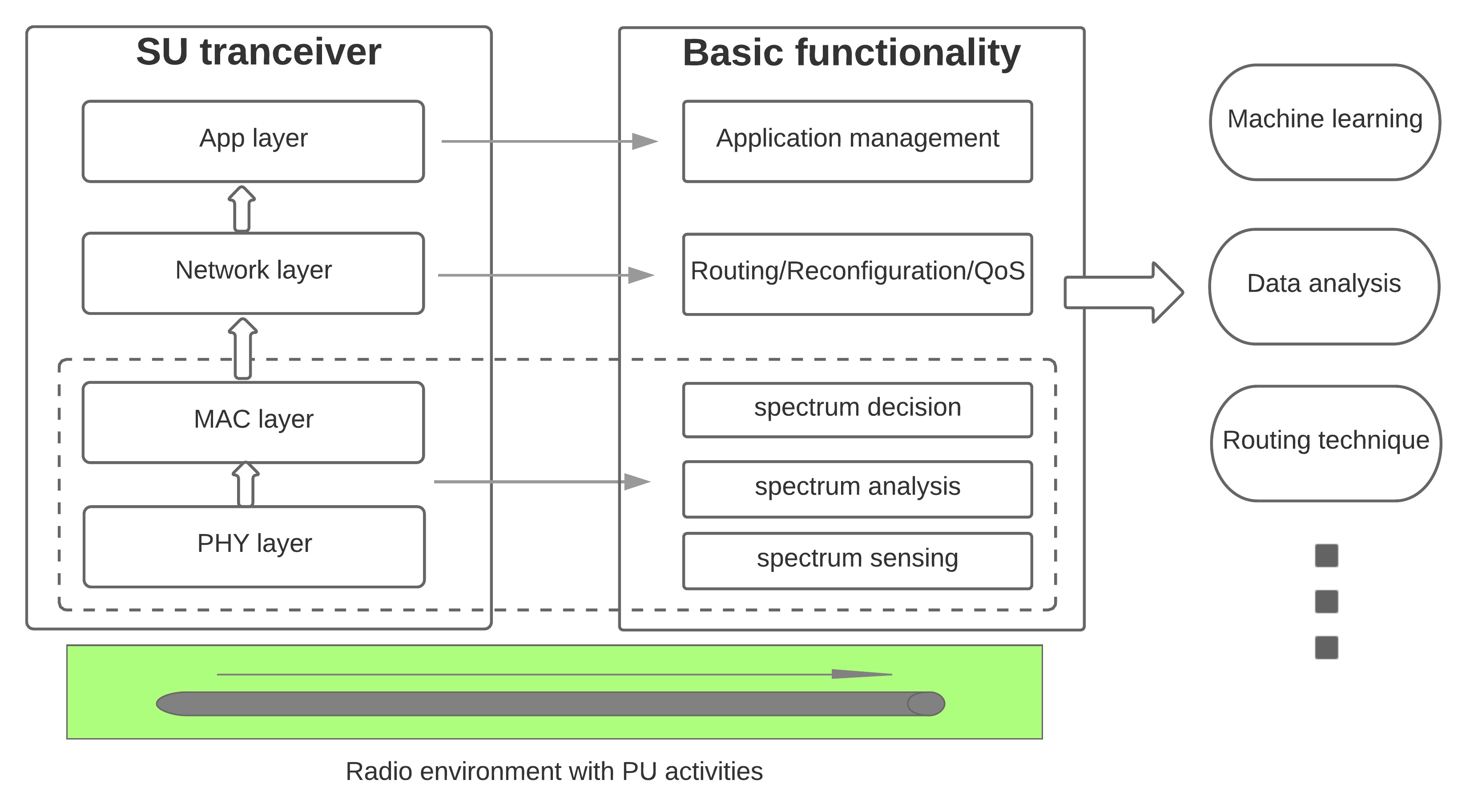}
	\caption{CRN tranceiver architecture.}
	\label{fig:tranc}
	\vspace{-15pt}
\end{figure}

The intrinsic feature of CRN is to find the best available channel allocation strategy for communication while maintain harmful-interference-free environment to PU network. Unlike traditional wireless networks, CRN is configured on a new basis of media access approach. Figure \ref{fig:tranc} shows a fundamental architecture of CR tranceiver. The tranceiver operates dynamic spectrum access (DSA) by sensing the spectrum, analyzing the spectrum and making joint decisions on spectrum selections. In high level functionalities, the tranceiver conduct routing operations, QoS monitoring, application management, as well as low level coordination and configurations. Thus, to model a full functional tranceiver should include these basic functions:

\begin{itemize}
\item receive spectrum waves from combined radio environment --- including PU activities, SU activities, and other noise.
\item spectrum sensing/analysis/decision ability --- detecting radio spectrum waves, extracting features, finding or predicting accessible spectrum slots by adopting data analysis and machine learning techniques, and making decisions on choosing spectrum slot for communication without introducing harmful interference to PUs.
\item global configuration/management/QoS control --- cooperating with other receivers and making joint decisions to meet minimal communication requirements and optimal spectrum resource allocation scheme.
\item routing --- conducting basic wireless network operations, sharing spectrum information and decision.
\item application management --- performing various applications.
\end{itemize}•

\subsection{Basic requirements for CRN simulators}
Despite well modeling of CR tranceivers, a well designed simulator for CRN experiment is expected to meet the following requirements:

\begin{enumerate}
	\item It is extensible with adaptable modular structures;
	\item It is easy for installation and configuration;
	\item It is able to simulate a versatile but controllable radio environment, such that illustrates how cognitive radio extents spectrum utilization efficiency;
	\item It consists of necessary signal process modules and data synthesis modules for further procedure of spectrum management;
    \item It includes the core feature of a CRN, an integrated network protocol stack structure for modification and reconfiguration; and
    \item It is preferred to be able to visualize the simulation process and result summarization for adjustment and modification.
\end{enumerate}

\subsection{Simulator Classification}
\begin{figure}[t]
	\centering
		\includegraphics[width=0.45\textwidth]{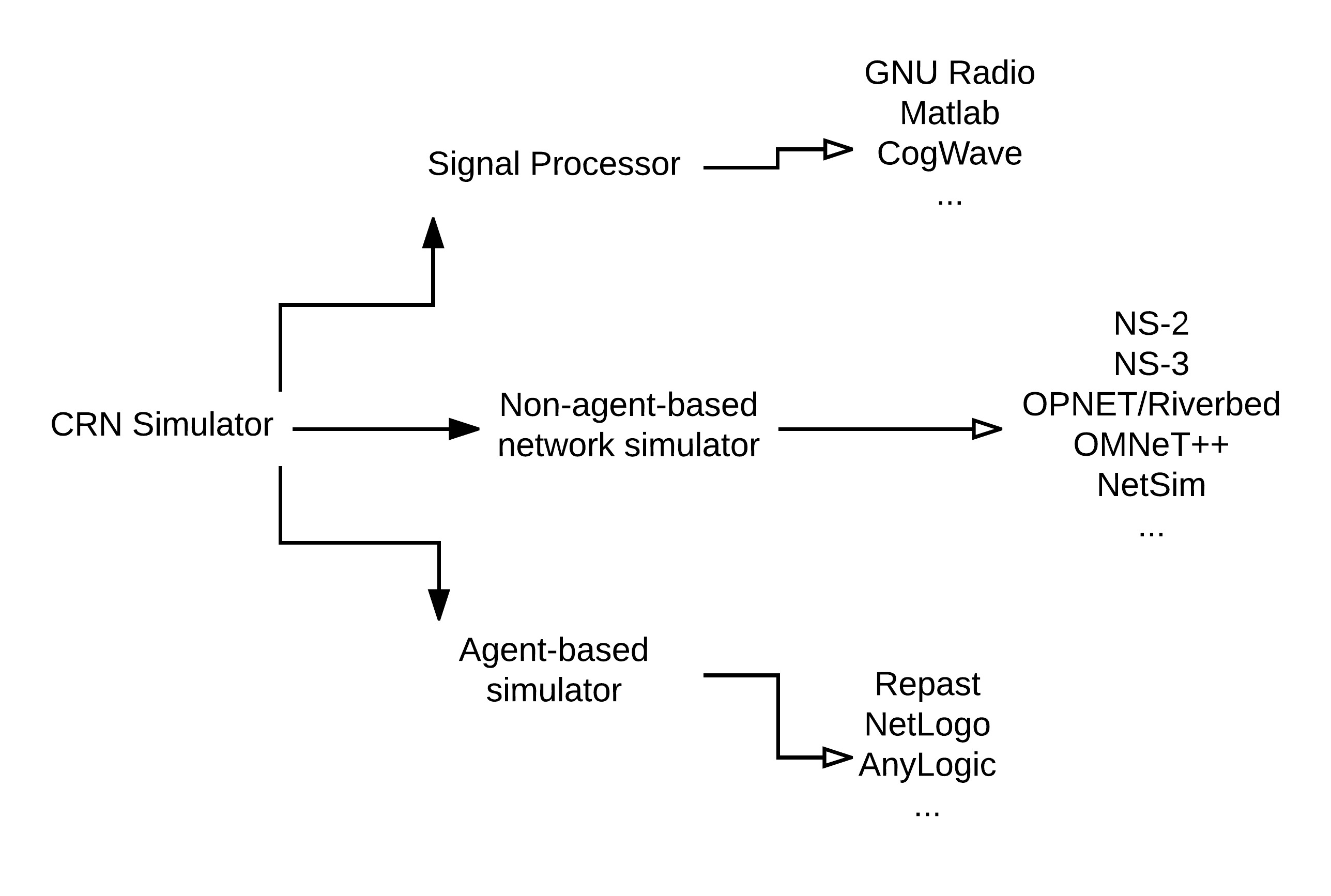}
	\caption{CRN Simulator Division.}
	\label{fig:simulator}
	\vspace{-15pt}
\end{figure}

Cognitive Radio study is complex and relates to many sub-fields. On physical level, waveform generation and analysis is critical; especially in early CRN studies, spectrum sensing is widely discussed, which requires high fidelity cognitive radio waveform. When considering the entire structure of CRN as a wireless network, comprehensive network simulator is desired for large-scale network simulation. In another prospective of CRN system, each cognitive radio can be regarded as an intelligent agent, thus agent-based modeling (ABM) is a well fitted concept to describe CRN simulation. Due to the long history of network simulation tools development, there are many of them can potentially be CRN simulation tools, some of them have even included customized packages for CRN simulation experiments. Under such provision, CRN simulator can be divided into three categories: signal processor, traditional non-agent-based network simulator, and agent-based simulator, as shown in Fig. \ref{fig:simulator}.

\section{Signal Processor}
\label{sec:radio}

\subsection{GNU Radio}
GNU Radio is a free and open-source software development toolkit that provides visualized signal processing blocks to study wireless communications and software radios \cite{GNURadio}. GNU Radio architecture is constructed by C++ and Python. With 15-year development, GNU Radio has a growing large signal processing block library, in which most of performance-critical modules are developed by C++ to reduce processing time of computational intensive blocks, while Python is used to develop non performance-critical blocks, and more importantly acts as glue to connect all signal processing blocks together and manages blocks control. GNU Radio provides a graphical user interface (GUI) and drag-and-drop manner to users to construct functional radios. Thus, GNU Radio is user-friendly especially for entree level researchers. In addition, it is flexible and comprehensive enough to customize block functions and connectivities by either modifying top level design or developing novel functional blocks. Along with great support and development from both community and official parties, GNU Radio becomes one of the most important tools on studying CRNs.

\begin{figure}[t]
\begin{minipage}[c][5.5cm][t]{.45\textwidth}
  \vspace*{\fill}
  \centering
  \includegraphics[width=7.5cm,height=2.5cm]{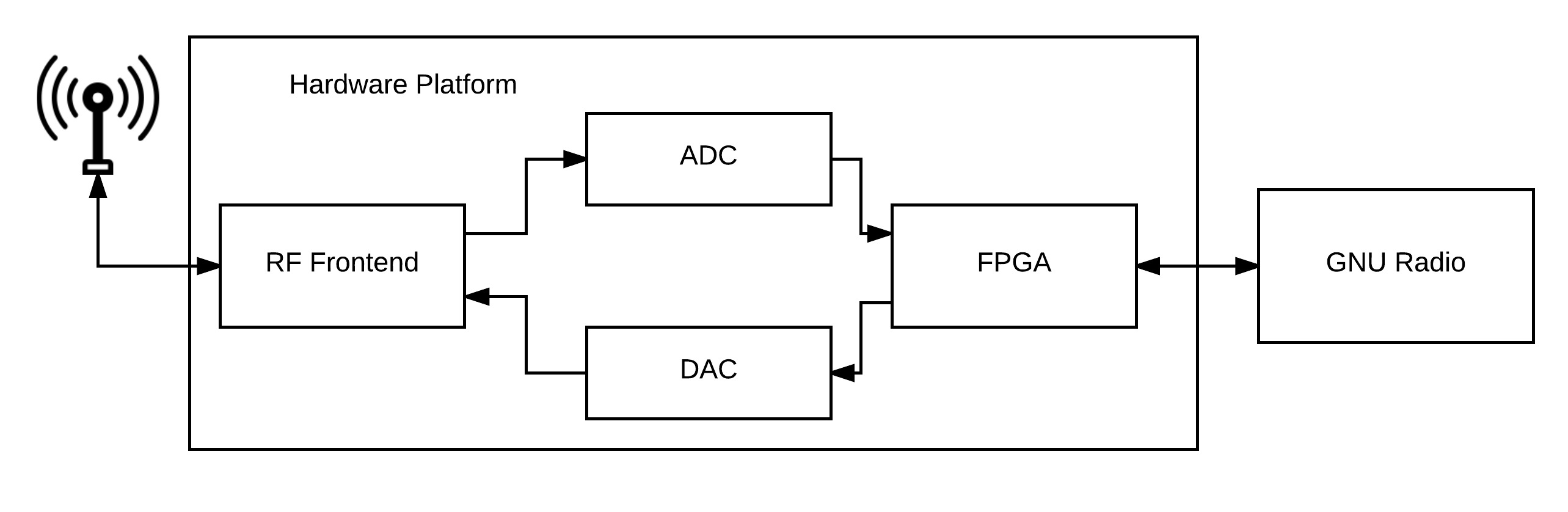}
  \subcaption{Diagram of simulation with hardwares.}
  \label{fig:hardware}\par\vfill
  \includegraphics[width=7.5cm,height=1.5cm]{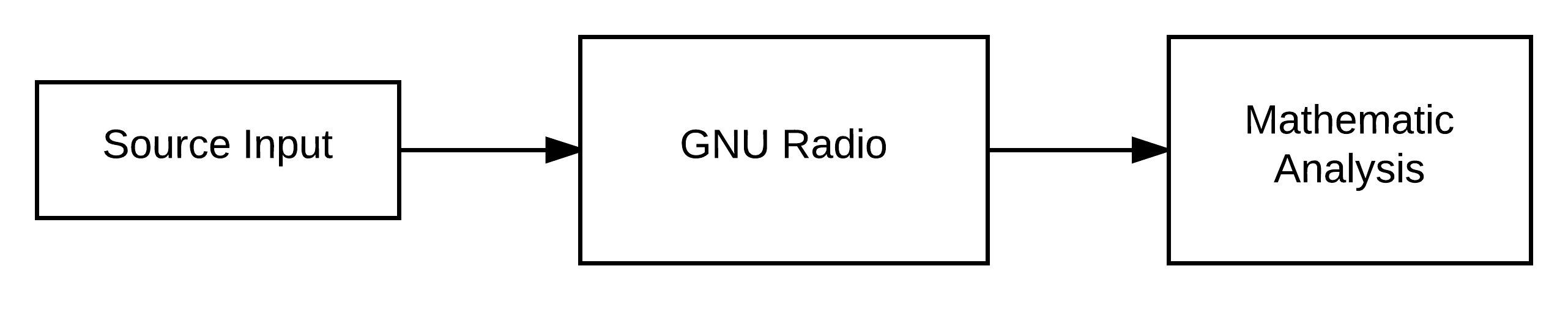}
  \subcaption{Diagram of simulation with other softwares.}
  \label{fig:software}
\end{minipage}
\caption{CRN simulation in GNU Radio.}
\end{figure}

One of the important feature of GNU Radio on CRN simulation, is that it is no more than a radio generator and signal processor. In CRN simulation, GNU Radio is usually used along with other tools, such as a radio tranceiver hardware platform (Figure \ref{fig:hardware}), or a mathematic analysis software (Figure \ref{fig:software}). GNU Radio is adopted for CRN-specified simulation from the early stage of CRN study \cite{Blossom:2004aa}. It embraces an amble library of signal processing blocks for CRN study in current version (v3.7.10.1), including: audio signal source, filters, OFDM block, other modulator and demodulators, digital tv blocks, channel models, etc. A great number of CRN studies used GNU Radio for experimental study \cite{Mate:2011aa}, \cite{Marwanto:2009aa}, \cite{Liu:2010aa}, \cite{Ding:2010aa}, \cite{Ouattara:2012aa}, \cite{Reyes:2015aa}, \cite{Rashid:2015aa}.

\subsection{Matlab}
As one of the most widely used scientific experimental tools with powerful mathematic toolboxes and packages, Matlab (Matrix Laboratory) is the pioneer on cognitive radio simulation because of its popularity on numerical computing and ease of use. In early research stages, Matlab is usually used to validate spectrum sensing and allocation schemes \cite{Maldonado:2005aa}, \cite{Wang:2008aa}. For example, an 802.11a/g OFDM WLAN PHY-layer communications system is established by introducing a dynamic spectrum access scheme in \cite{Maldonado:2005aa}, which enhanced the performance of network capacity and throughput. In \cite{Wang:2008aa}, a CRN is embedded in a multiple primary radio exist environment with a specific signal propagation model. In this complex communication environment, a high energy efficiency, high throughput spectrum sharing algorithm is tested.

Matlab is ideal for CRN PHY layer simulation, because it is naturally deployed to process signals, build up transceiver model, and further to set up communication systems. With proper system support in Matlab, researchers have implemented multiple detection schemes such as energy detection, matched filter detection, cyclostationary feature detection, cooperative detection and interference-based detection \cite{Ghosh:2014aa}. In \cite{Kondareddy:2008aa}, a novel broadcasting scheme of CRN was inspected in which 100 nodes were randomly scattered in an $1000m \times 1000m$ area with limited spectrum resource for broadcasting.

\begin{figure}[t]
	\centering
		\includegraphics[width=0.45\textwidth]{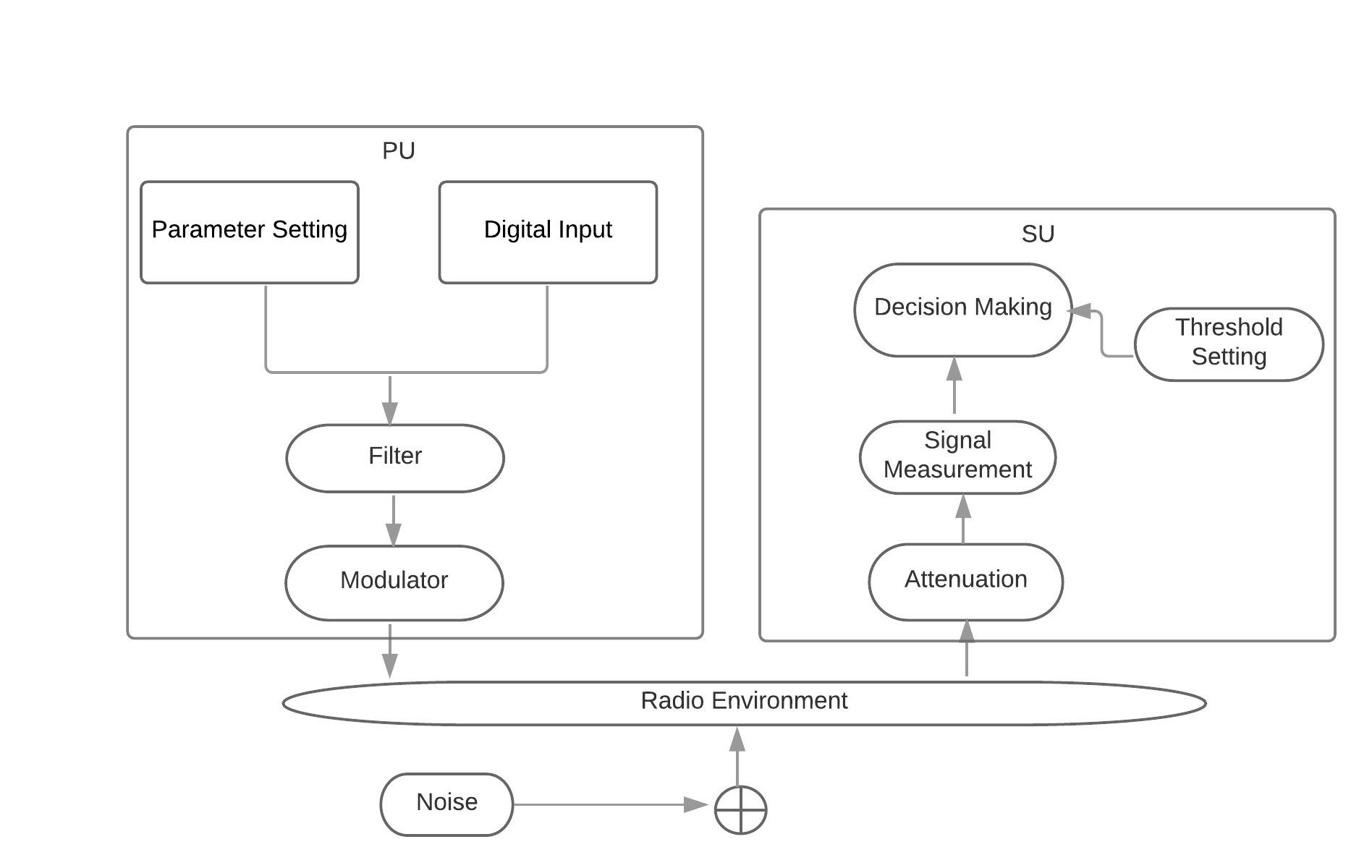}
	\caption{Simulation diagram of CRN energy detection in Matlab.}
	\label{fig:mat}
	\vspace{-15pt}
\end{figure}

A case example of simulating CRN functionality in Matlab is shown in Fig. \ref{fig:mat}. Energy detection is the most fundamental technique on spectrum sensing in CRNs. To justify the feasibility and viability of energy detection, PU signal is first generated with added noise in radio environment. With proper radio propagation model, the received signal can be simulated in SU side, and energy detection technique is easily adopted by measuring received signal level.

In addition, Matlab has been utilized to simulate cognitive radios from many other aspects. MAC layer protocols have been studied using Matlab in recent research \cite{Hu:2012aa}. For instance, a countermeasure against the most active band attack was simulated using Matlab in \cite{Hu:2012aa}. Also, Matlab was used to study Denial of Service (DoS) attacks in CRNs. An intrusion detection system (IDS) against various types of CRN attacks has been studied in \cite{Fadlullah:2013aa}. In this work, 68 available communication channels were distributed in broadcasting TV channels with 12 secondary users, and other network parameters were set according to real situation configurations. By testing detection latency under different attack scenarios, the IDS was proven efficient on against DoS attacks. As a matter of fact, Matlab is more than a signal processor; it is a comprehensive experiment tool can be widely used in CRN study.

With plethora of toolboxes, especially supported by Simulink, Matlab can specify design details of network component as well as topology, which makes it ideal for operation on signal processing and spectrum management. Usually, Simulink is cooperatively used to drive CRN hardware platforms for experimental analysis. In \cite{Tkachenko:2006aa}, Simulink is used to construct a CR front-end and then translated to drive BEE2 enhanced Xilinx platform. Similarly, a CRN experiment example is given in \cite{Anas:2012aa} that Simulink is jointly used with USRP platform on spectrum sensing.

However, due to lacking of pre-defined network architecture, Matlab/Simulink is not convenient for higher level network simulation or comprehensive construction studies.

\subsection{Discussion}

GNU Radio and Matlab share many similarities, such that both of them support for customized functional modules, great on signal processing, ease-of-use on drag-and-drop manner, real-time monitoring on signals. Another very important feature of them, is they all developed with APIs to some famous signal processing hardware platforms, such as Universal Software Radio Peripheral (USRP) \cite{USRP} developed by Ettus Research for CRN study.

Matlab is not simply a signal processor. Instead, it is a large-scale software with radio generation and network simulation capability. Thanks to its rich mathematic library and great computational power, Matlab is widely used in study CRN from signal analysis of spectrum sensing, partially network modeling, cognitive radio generation, etc. In comparison, GNU radio is more like an pure radio generator and signal processor with great extension to using hardware radio platforms for real-environment CRN emulation. However, those signal processor is limited in CRN simulation by lacking of comprehensive network structure, thus for some intensive studies of CRNs, they are usually not a popular choice.

\section{Non-Agent-based Network Simulator}
\label{sec:non}
Most traditional network simulators are non-agent-based simulators. Existing non-agent-based simulation packages fall into two categories: general network simulators with additional wireless networking components such as NS-2, OMNET++, and wireless network specified simulators such as J-Sim \cite{dzikowski2009agent}. This subsection will introduce some of them and illustrate how each of them being used in different CRN research works. At last, their advantages and constraints will be highlighted.

\subsection{NS-2}

Network Simulator 2 (NS-2) is a general purposed network simulator that has been widely used in computer network research community for years \cite{issariyakul2011introduction}. This section introduces the characteristics of NS-2 in high level first, then some applications of NS-2 in CRN study are discussed.


NS-2 is an object oriented simulator. Its framework is developed by C++, and front-end is interpreted by OTcl. With access to very detail network modules designed for modification, NS-2 is usually adopted for protocol design, network performance comparison, network traffic study, and new architecture design, including range from physical layer to network layer or even higher. Thus, NS-2 is often extended to study new protocols or packages. Since NS-2 is friendly to embrace new development, it is naturally developed for CRNs. NS-2 has many unique features for CRN simulation. Below is a concise list of its advantages and constraints.

\newcolumntype{P}[1]{>{\centering\arraybackslash}m{#1}}
\begin{table*}[t]
\begin{center}
\caption{Comparison of different signal processors.}
\label{tab:sim}
\begin{tabular}{|P{2.3cm}|P{3.5cm}|P{2.3cm}|P{1.8cm}|P{3.0cm}|}
\hline
 Simulator &  License & APIs to CRN hardware platforms&  Platform & Works\\
\hline
GNU Radio & Open source&Yes & Linux/Limited functionality on Windows & \cite{Mate:2011aa} \cite{Marwanto:2009aa} \cite{Liu:2010aa} \cite{Ding:2010aa} \cite{Ouattara:2012aa} \cite{Reyes:2015aa} \cite{Rashid:2015aa}\\
\hline
MATLAB/Simulink & Commercial& Yes & Multiple & \cite{Maldonado:2005aa} \cite{Wang:2008aa} \cite{Maldonado:2005aa} \cite{Wang:2008aa} \cite{Ghosh:2014aa} \cite{Kondareddy:2008aa} \cite{Hu:2012aa} \cite{Hu:2012aa} \cite{Fadlullah:2013aa} \cite{Anas:2012aa} 	\\
\hline
\end{tabular}•
\end{center}
\end{table*}•

\textit{Advantages}:
\begin{itemize}
  \item NS-2 has a large number of available network architectures and models that can be easily accessed and customized for specific simulation tasks.
  \item NS-2 supports simulation of wireless networks with many embedded protocols and modules, such as mobility module, energy module and propagation module.
  \item NS-2 has support from large group of users, as it is under on-going development with expected generation of cognitive radio modules.
  \item NS-2 possesses good flexibility that allows users customize the format and content of output trace file. 
  \item It is easy to adjust the topology of the network by modifying front-end script files.
  \item NS-2 is a discrete event scheduler, such that all events in network simulation are controllable.
\end{itemize}

\textit{Constraints}:
\begin{itemize}
  \item NS-2 is a complex system contains many function blocks. And it is constructed using two languages C++ and TCL, which makes the structure difficult to learn for beginners.
  \item NS-2 requires recompilation every time if there is an update on system structures (not including network construction script files).
  \item NS-2 adopted TCL for network simulation, which brings large overhead of execution time due to the TCL interpreter.
\end{itemize}

\begin{figure}[t]
	\centering
		\includegraphics[width=0.45\textwidth]{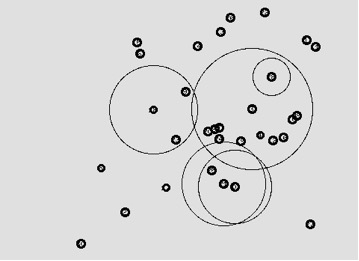}
	\caption{Visualization of NS-2 simulation (Nam).}
	\label{fig:nam}
	\vspace{-10pt}
\end{figure}

The simulation process can be visualized by using Nam as shown in Fig. \ref{fig:nam}, which is simply a optional demonstration of simulation process. Actually, it is more efficient if we turn off the visualization functionality. In addition, it is convenient to choose the output trace file format and content. Then, the user can focus on result analysis using different data analysis tools.

A new MAC protocol named AMAC that aims to improve network throughput was simulated using NS-2 \cite{joshi2009efficient}. In \cite{kondareddy2008}, a plain 80-node network topology has been constructed with half of the nodes for transmission and another half for receiving. In this network environment, two MAC protocols were compared in terms of throughput, where there are three channels available for communication. In \cite{joshi2011enhanced}, several scenarios have been studied to testify the performance of a synchronized MAC protocol.

In addition, NS-2 has also been used to testify the efficiency of security schemes against to malicious attacks. In \cite{wu2011improved}, the authors simulated a CRN with 60 normal nodes and 40 selfish nodes using TCP as transport protocol for transmitting 1Mb/s CBR data stream, where selfish nodes sent false information to control channel following the Poisson process. The authors of \cite{zhu2008two} simulated cognitive radio attack with implementing 802.11 (WiFi) protocol, and CSMA/CA wireless medium access scheme.

Beside the general protocol stacks in NS-2, there are other CRN specified packages such as CRCN (Cognitive Radio Cognitive Networks) simulator (download from \cite{crcn_url}), which can be integrated into NS-2. CRCN has been developed as an important supplement to CRN simulation. It supports development of cognitive radio protocols for both MAC layer management and network layer routing. Generally, it mainly includes two newly developed MAC layer protocols: MACCON and MACNG. Both of them are not complex; MACCON simply use ``aloha'' strategy on tranceiving packets, while MACNG will evaluate channel status prior to tranceiving. The CRCN patch is useful on evaluating performance of basic CRN functions with limited directly use. But it is well structured for further CRN development, including energy management, dynamic spectrum allocation, spectrum sensing, and security management. Although it is a great add-up package can be used for cognitive radio simulation, CRCN has been developed as an quite isolated package and is only compatible with an older version of NS-2 (version 2.31), which might cause restrictions on its extensive use in academic study.

In \cite{rahmani2014performance}, the authors comprehensively analyzed several MAC protocols developed for cognitive radio applications, including MAC-802.11, MAC-Simple, MACCON and MACNG. Besides, the author compared the performance from variant perspectives, such as throughout, end to end delay, and packet delivery rate for each MAC protocol when adopted three different upper layer routing protocols: AODV, DSDV and DSR.

The integrated open source protocols in CRCN allows researchers to easily simulate CRN with various protocols and topologies by simply revising the script. In \cite{saadat2013mobility}, multiple propagation models, such as free pace propagation model, two ray ground propagation model and shadowing propagation model, have been compared from multiple aspects in CRNs, leveraging the support of CRCN patch and MPEG4 patch. These reported work has shown that CRCN integrated many features and protocols of cognitive radio into NS-2, which sheds a light on simulating CRN, but still needs further development.

Meanwhile, many other NS-2 patches to for CRN function blocks, such as CogNS \cite{Esmaeelzadeh:2013aa} and CRAHNs \cite{Di-Felice:2011aa} are under development. It is foreseeable that more efforts will be reported in the coming years.

\subsection{NS-3}
Network simulator 3 is an open source, discrete-event driven, network simulator targeted primarily for research and educational use. As a successor of the NS-2, NS-3 is entirely written in C++ that enables a simulation to be developed purely in C++ to reduce compilation time. It is not applicable in the NS-2 because it is necessary to run OTcl script in NS-2. In NS-3, Python programming language is optional in scripting and visualization. Although NS-2 and NS-3 have many similarities such as framework structure, style, protocol stacks, application library, and simulator support, comparing to NS-2, NS-3 has following advantages \cite{Henderson:2008aa}:

\begin{enumerate}
   \item It improved scalability and modularity with a newly designed software core;
   \item It is more realistic to reflect features of network nodes;
   \item It is more software integrated;
   \item It supports virtualization to run virtual machines over simulating; and
   \item It can be adapted to work in real devices such as testbeds for research-oriented tasks.
\end{enumerate}

Thanks to these advanced features, NS-3 is more suitable for large scale simulations including large scaled CRNs \cite{Al-Ali:2014aa}. In \cite{Bicen:2012aa}, a cognitive radio sensor network is simulated by extending the NS-3 network simulator, in which 200 nodes and a sink are scattered in a $100m\times 100m$ field with existence of 10 primary users. Network performance has been elaborately studied in different environments.

\begin{figure}[t]
	\centering
		\includegraphics[width=0.45\textwidth]{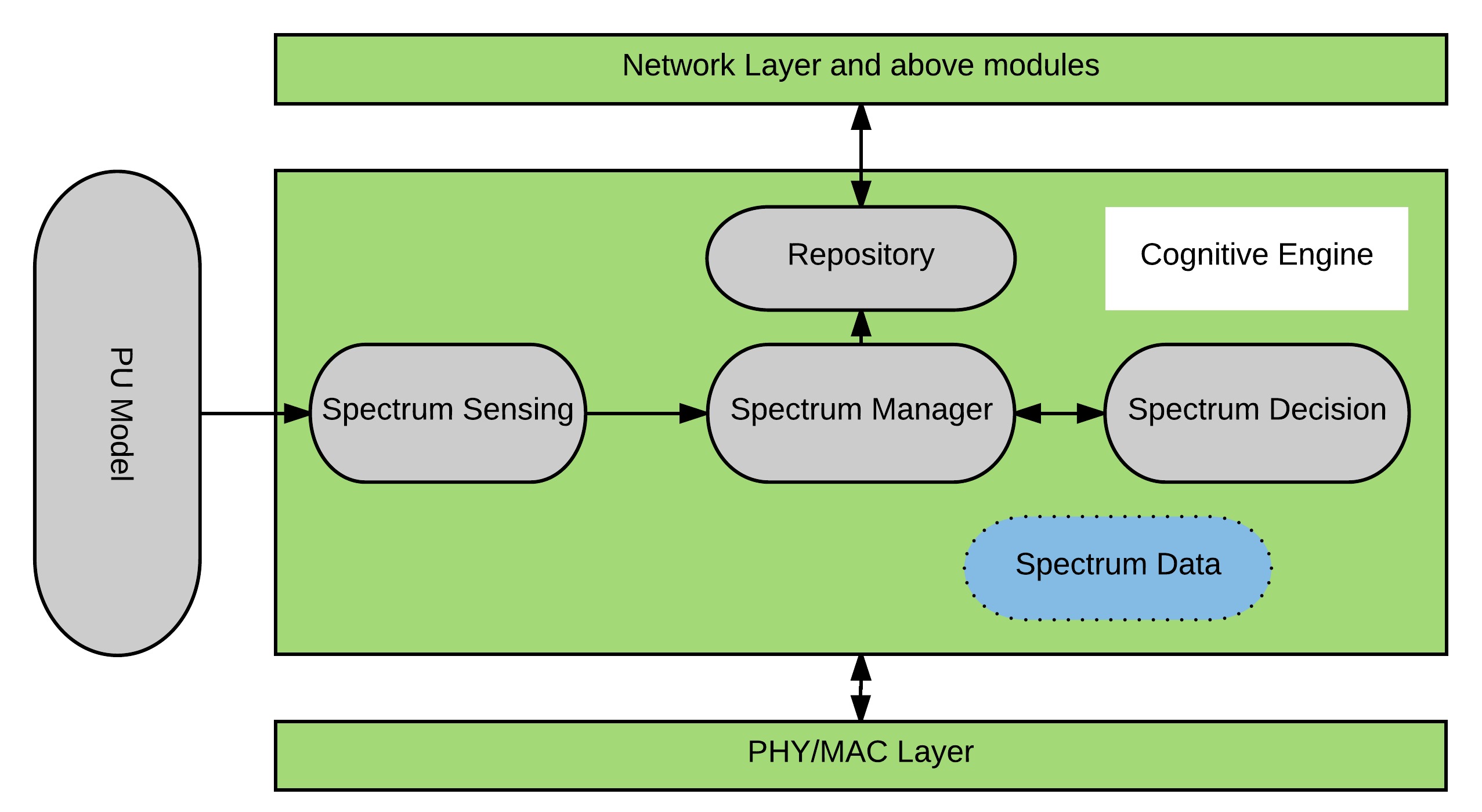}
	\caption{The block structure of CRE-NS3 extension in ns-3.17.}
	\label{fig:cre}
	\vspace{-18pt}
\end{figure}

Recently, a new upgraded version of NS-3 is released to include CRN functionalities, called CRE-NS3 (Cognitive radio extension for ns-3, github repository can be found at \cite{ns3_url}). In 2014, A. Al-Ali and K. Chowdhury have proposed the prototype of CRE-NS3 extension \cite{Al-Ali:2014aa}. In this work, substantial effort has been undertaken on modifying transport layer, network layer, link layer and physical layer by changing packet structure, defining new concept of cognitive interface, adding channel processing state, and introducing new protocols. The simulation performance of NS-2 and NS-3 is also compared in \cite{Al-Ali:2014aa}. The result indicates that NS-3 outperforms NS-2 by far less memory usage and slightly less execution time. Essentially, CRE-NS3 extension is mainly an extended layer between PHY/MAC layer and network layer as shown in Figure \ref{fig:cre}. It is concrete to include full stack of CRN modules in spite of simple realizations. In addition, CRE-NS3 gains great contribution from many developers via either github repository \cite{ns3_url} or public web-page \cite{ns3web}.

\subsection{OPNET/Riverbed}
Optimized Network Engineering Tool (OPNET) is a widely used object oriented, discrete event driven, general purpose network simulation tool with many user preferred features, tool sets and a huge library of available network protocols. It helps set up different network topologies and is convenient for users who need to design customized network protocols or applications. Generally speaking, using OPNET for experimental study consists of three main procedures: modeling, simulating and analysis. In simulation, all required models of protocols should be created and defined first. Then, the network is simulated based on the developed models with hierarchical structures. OPNET defines a network as a collection of sub-models, in which the hierarchical structure is divided into three domains:

\begin{enumerate}
   \item The network domain: the top level. It defines the high level information of the system such as network topology, overall configuration, etc.
   \item The node domain: defines the internal structure of network nodes. A node is constructed by integrating many different functional modules.
   \item The process domain: the lowest level where the user programs module functionalities including routing protocols, communication algorithms, memory utility, etc.
\end{enumerate}

OPNET provides integrated result analysis tools that graphically displays the selected simulation results in scalar graphs. As a commercial supported software, OPNET has powerful graphical support of network topology, traffic flow control and various entities of network units. OPNET has a large user group and many discussion forums, so it is easy to be extended to implement CRNs. In October 2012, the Riverbed company announced that OPNET has become part of Riverbed Modeler.

OPNET provides many integrated propagation models such as Free Space, Hata, CCIR, Longley-Rice, TIREM, Walfish-Ikegami, and more added models such as Rayleigh, Ricean, Two-Ray, and 802.15.4/ZigBee models from OPNET community \cite{Timm-Giel:2008aa}. These models enable OPNET to accurately model the radio transmission. In \cite{Jafri:2011aa}, the authors tested the performance of a small network with 10 PU nodes and 4 SU nodes. With specified definition of node structure, overall transmission configuration, packet format and a new MAC protocol, the simulation result was examined by detailed graphical result analysis. In \cite{Faruk:2012aa} and \cite{Faruk:2013aa}, two improved MAC protocols based on CSMA/CA MAC protocol and the IEEE 802.11b were measured using OPNET, both of which extended the use of white space of spectrum. In \cite{Thomas:2012aa}, a multi-hop cellular network working with CR technique was implemented by OPNET. A multi-interface CR mobile node model was created for communication in multiple ways. Then, a spectrum sensing and management protocol and a high-level routing protocol were implemented upon the model. The experiment was simulated with many adjustment of different parameters.

OPNET/Riverbed modeler is powerful with great document and technique support due to it is a commercial simulator. It is very suitable to simulate network behaviors in the real world as including real world scalers and maps. For the same reason, however, it is not popular as an academic CRN simulator because of its high expense.

\subsection{OMNeT++}
The Objective Modular Network Testbed in C++ (OMNeT++) is a well-designed open source, component-based, discrete event driven simulator written in C++. As an open source network simulator, OMNeT++ have plenty of public documents like NS-2 and NS-3 do, and it is also equiped with a powerful graphical user interface like OPNET does. Thus, OMNeT++ becomes successful in both industry as well as academia.

OMNeT++ provides a hierarchically nested architecture for modeling. An OMNeT++ model is composed by many modules that pass messages back and forth for communication. Simple modules are the basic active elements used to model a system. They are programmed in C++ using the OMNeT++ simulation class library. Users can design new simple modules in OMNeT++ IDE. Meanwhile, a large number of simple modules are provided in the OMNeT++ library. Single modules can be assembled into more functional compound modules and so forth for compound modules with unlimited hierarchy levels.

\begin{figure}[t]
	\centering
		\includegraphics[width=0.45\textwidth]{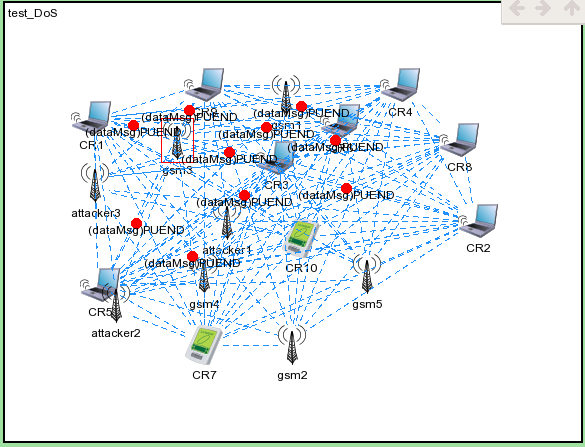}
	\caption{An OMNeT++ GUI example of CRN simulation.}
	\label{fig:omnet}
	\vspace{-12pt}
\end{figure}

NED is the adhesive language in OMNeT++ for declaring simple modules, assembling for compound modules and constructing network topologies. In the process of simulation and after, Tkenv, a integrated GUI of OMNeT++, provides users powerful features for on-going animation tracing, output tracing and synthesizing. It is convenient for user to look for output result of the system or even each modules. Figure \ref{fig:omnet} shows an example of CRN simulation by OMNeT++ simulator.

On top of OMNeT++, MiXiM (Mixed Simulator) has been developed as a simulation framework for wireless and mobile networks. It provides wireless MAC protocols, as well as detailed models of radio wave propagation, interface estimation, and radio transceiver power consumption. In MAC layer, IEEE 802.11 and IEEE 802.15.4 protocols have been developed. Due to the component-based structure, it is relatively less time consuming to implement a new CRN protocol via OMNeT++.

To study cognitive radio, there are some reported efforts that have built up proper framework and simulation models using OMNeT++, which shedded lights on developing CRNs efficiently. Specifically, J. Marinho and E. Monteiro implemented a CR framework with detailed modules design, construction pattern, and code style upon OMNeT++/MiXiM \cite{Marinho:2011aa}. The authors have validated two MAC protocols as extensions of the base CR MAC module in the proposed CR framework.

\begin{figure}[t]
	\centering
		\includegraphics[width=0.45\textwidth]{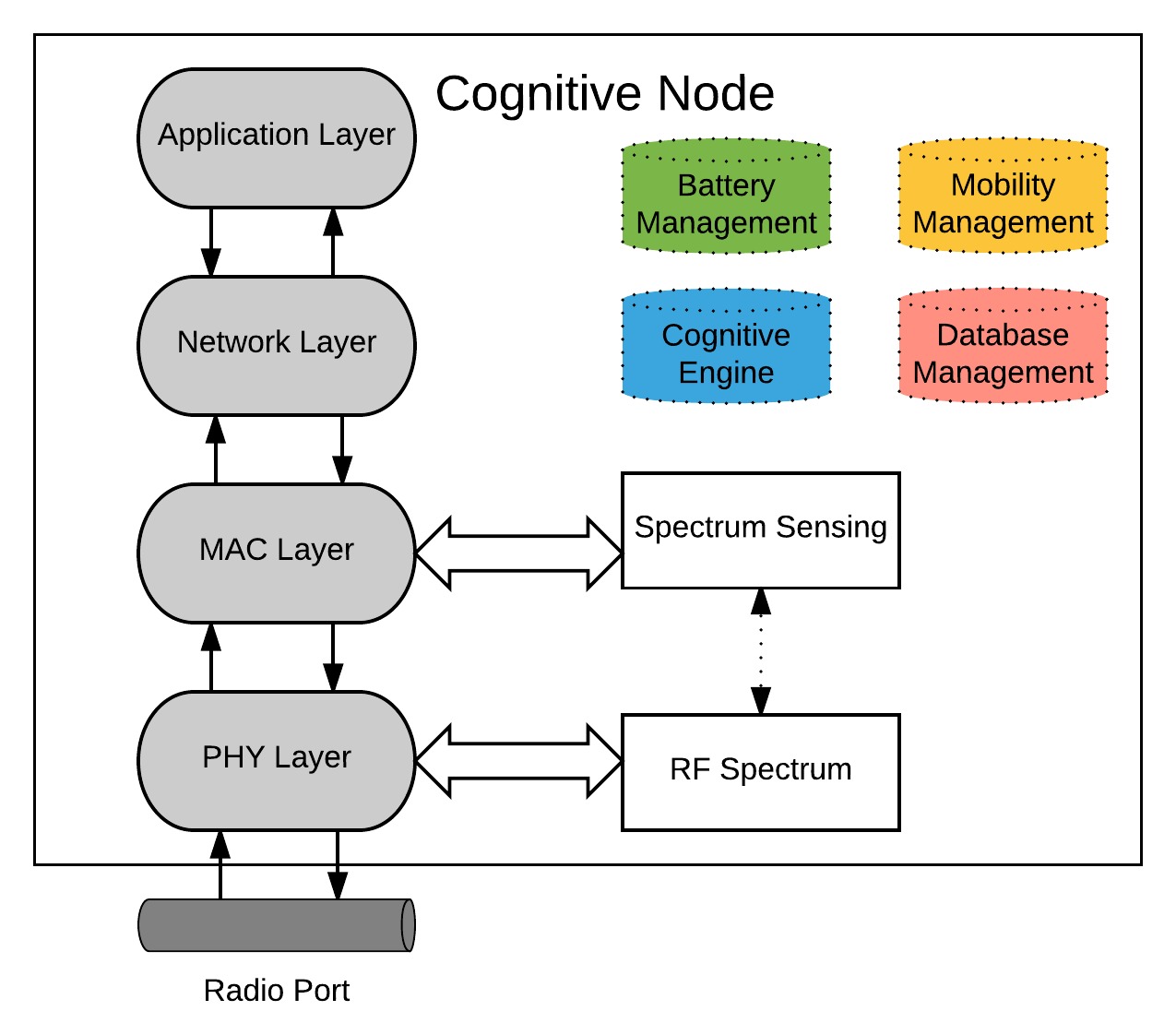}
	\caption{The node structure of crSimulator in OMNeT++.}
	\label{fig:crsimulator}
	\vspace{-12pt}
\end{figure}

Another important cognitive framework named crSimulator model has been built by Shah Nawaz Khan and others in \cite{Khan:2013aa}. This crSimulator framework mainly consist of three functional layers: application layer to generate various data flows, CR MAC layer to simulate different MAC protocols, and physical layer to simulate transceiver and particular transmission channels. It is a fully developed stack of CRN simulator package. The node structure introduced by crSimulator is shown in Figure \ref{fig:crsimulator}. The cognitive node in crSimulator is constructed layer by layer with various spectrum sensing capabilities, and other capabilities as a wireless/cognitive node, such as battery management, mobility management, database management and cognitive engine.

OMNeT++ is also used to simulate different MAC protocols in different environment by adjusting several modules and structure composition. In \cite{Sebastien:2011aa}, a spectrum allocation scheme is simulated and tested in high speed mobility environment (vehicle traveling) in OMNeT++/MiXiM. In \cite{Ahmed:2013aa}, another spectrum allocation scheme is simulated in an 1000x1000 $m^2$ area with fixed number of SUs and varied number of PUs. In other works such as \cite{Quach:2012aa} and \cite{Quach:2013aa}, OMNeT++ has been used to simulate CR performance in radio overlap regions with specified IEEE 802.11 protocol.

In summary, OMNeT++ is a simulator that has been used in many aspects in CRNs such as inspecting spectrum management schemes, implementing MAC protocols, and analyzing radio interplay interference in intricate environment \cite{Ahmed:2014aa}, \cite{Iyer:2011aa}. Because of the easiness and open source, OMNeT++ is one of the most popular simulator for CRNs.

\subsection{NetSim}
The NetSim is a comprehensive, stochastic discrete event driven computer network simulator that can simulate various network hardwares and softwares. It is usually used in studying Cisco network and training people for network experience. NetSim was born with a built-in development environment with excellent GUI support. Users can benefit from a simple drag-and-drop pattern of network construction process. It comes with abundant protocol libraries and models including many wireless supports such as WLAN, IEEE 802.11 a/b/g/n, GSM, CDMA, Wi-Max, MANET,  Wireless Sensor Network, and Zigbee. Furthermore, as a commercial product, NetSim is one of the few network simulators with integrated cognitive radio module. It supports direct simulations on CRNs, particularly for simple applications. Protocols in NetSim are developed with open C code. Users can modify existing protocols as well to build their own.

It is noteworthy that NetSim is more compatible with Windows system than Linux system, and it is more friendly to Visual Studio IDE. Thus, for those who do not use Windows system, they have to run a Windows virtual machine or Windows enabled platform on Linux system to enable NetSim for simulation, which slightly limits its use in academic studies. On the other hand, with support of Visual Studio, it is very user friendly for debugging and testing. Despite of this limitation, NetSim has gained popularity on simulating cognitive radio as either teaching instrument or research tool. For example, in \cite{Shalini:2014aa}, NetSim was used to simulate a small scale CRN using the integrated cognitive radio model. In \cite{Kanti2015}, several modulation schemes are compared in CRN with five cognitive radios and two PUs.

Comparing to other traditional network simulators such as NS-2 or OMNeT++, NetSim is less popular in cognitive radio simulation. But, as a powerful and user-friend simulator, we believe that NetSim will be adopted by more users in the community for CRNs studies in the future.

\newcolumntype{P}[1]{>{\centering\arraybackslash}m{#1}}
\begin{table*}[t]
\begin{center}
\caption{Comparison of different simulators.}
\label{tab:sim}
\begin{tabular}{|P{1.8cm}|P{1.6cm}|P{2.2cm}|P{2.3cm}|P{1.8cm}|P{3.0cm}|}
\hline
 Simulator & Interface &Cognitve Framework  & License &  Platform & Works\\
\hline
NS2	& C++/OTCL & CRCN/CRAHN  &   Free & Linux &\cite{joshi2009efficient} \cite{kondareddy2008} \cite{joshi2011enhanced} \cite{zhu2008two} \cite{rahmani2014performance} \cite{saadat2013mobility} 	\\
\hline
NS3	& C++/Python & CRE-NS3 \cite{Al-Ali:2014aa} &  Free & Linux&\cite{Bicen:2012aa} \cite{Chang2013} \cite{Mao2015} \cite{Chang2015}\\
\hline
OPNET/Riverbed  & C/C++ & No  &  Heritage OPNET is Academic free & Multiple &\cite{Jafri:2011aa} \cite{Faruk:2012aa} \cite{Faruk:2013aa} \cite{Thomas:2012aa} \cite{Hossain2012} \cite{Abdelraheem2015} \cite{Bri2015} \\
\hline
OMNeT++ & C++/NED & crSimulator\cite{Khan:2013aa} and others  &  Free &  Multiple& \cite{Sebastien:2011aa} \cite{Ahmed:2013aa} \cite{Quach:2012aa} \cite{Quach:2013aa} \cite{Ahmed:2014aa} \cite{Iyer:2011aa} \cite{Li2016} \cite{Trotta2015} \cite{Tushir2016}\\
\hline
NetSim &	C & 802.22 &  Academic free  & Windows&\cite{Shalini:2014aa} \cite{Kanti2015}  \cite{Rath2016}	\\
\hline
\end{tabular}•
\end{center}
\end{table*}•

\subsection{Others} 

There are other network simulators that can be used for cognitive radio simulation, including J-Sim, SENS, ATEMN, QualNet, GloMoSim, and TOSSIM. Although they are not considered as very popular simulators, they are potentially to be suitable for cognitive radio study \cite{Mehta:2010aa}.

\subsection{Comparison and Analysis}
In the long history of computer network simulator development, different simulators were equipped with different features of certain advantages and constraints on simulating CRNs. These features vary in many aspects: length of development cycle, maturity, popularity, application basis, technique support, open source, scalability, ease of use, and integrated content.

In computer network research, users follow their own demands of design to select simulators. For those who only study top level network and verify common cognitive modules, it is natural to pay more attention on protocol library and easiness of simulation. On the other hand, for those who focus on developing new cognitive radio architectures, high flexibility of design pattern is required. Notably, OPNET, OMNeT++ and NetSim are featured with excellent graphic user interface that allows user to obtain good visualization of network component and topology in simulation. In comparison, NS-2 and NS-3, because of high complexity, are less user friendly on top level development and entry level design. While considering flexibility of design, NS-2 and NS-3 are more competent to satisfying users' flavor.

Based on our observation, open source and length of development cycle are two other critical factors that affect the popularity among users. Considering above discussed simulators, most of them are open source and free of charge. Others like OPNET and NetSim have academic versions, which are free of charge for academic use but with limited functionalities. Comparatively speaking, NS-2 and OMNeT++ gain more popularity in general computer network simulation. They have attracted more attention from CRN research community too, because there are plethora of materials and documents that serve as introductions to those who are new to the research work. That is also one of the main reasons that NS-2 are more popular than NS-3, although NS-3 is more efficient as a network simulator. Undeniably, NS-3, on the other hand, is growing its popularity gradually because of its great support and updates from its open source community.

Another reason for their popularity is that they are developed with rich library of protocols and models along with their widely use. As the most mentioned simulator, NS-2 lacks many infrastructure components that OMNeT++ and OPNET provides, such as hierarchical design pattern, GUI support for simulation process and topology editor, separation of models from experiments, graphical analysis tools, integrated analysis tools, etc. However, NS-2 is still superior in developing particular cognitive radio models.

OPNET and OMNeT++ are quite similar in many aspects except their evolution pattern. Apparently, OPNET provides more technical supports as a commercial product. It is also equipped with a larger protocol model library. In other sense, however, OMNeT++ is more suitable for new protocol design because of its open source nature.  It is worth to note that NetSim has inherited many good features from OPNET and OMNeT++. It is expected to be more popular in the near future.

In another sense of CRN simulation, non-agent-based network simulators are not quite suitable for spectrum sensing and spectrum analysis study, because they are naturally lack of signal processing modules and not focus on spectrum level realization. As a result, many CRN simulations are not merely rely on a single traditional network simulator; a combination use of signal processor and network simulator is welcome on designing full stack of high fidelity CRN simulations \cite{LeNir2015}.

Table \ref{tab:sim} summarizes the discussion of this subsection.

\section{Agent-based Simulator}
\label{sec:agent}
Agent-based simulators model the systems with components of autonomous, interacting agents. An agent is an identifiable individual with well-defined boundaries and a set of rules for behaviors. It is embedded in certain network to receive inputs from surrounding environment, intelligently interact with other agents, learn to adapt its environment and delegated to achieve the predefined goals \cite{macal2008agent}.

In CRNs research, agent-based simulation allows researchers more focus on some particular behaviors and rules instead of being overwhelmed by the complex system level behaviors. In some work \cite{Dzikowski2015}, cognitive radio is envisioned as an agent, thus CRN is regarded as a multi-agent system. In CRN simulation by agent-based simulator, the modeling process usually includes 1). defining behavior rules of each individual agent, and 2). assign all the agents into the simulated environment. Although the agent-based simulators are not as popular as traditional simulators in CRN simulation, they can be useful when considering ``cognition'' as someone's learning process in some environment. Due to the feature of flexibility in system configuration and re-usability of developed function modules, it is noteworthy to build CRN module in agent-based simulator.

\subsection{Repast}
Recursive Porous Agent Simulation Toolkit (Repast) is an open source software framework initially developed by researchers from the University of Chicago \cite{Collier:2003aa}. It is written in Java and can run in any systems supporting JVM. As one of the most popular agent-based modeling (ABM) software, Repast features with excellent GUI support for ease of use. It is programmable in many IDEs such as NetBeans, Eclipse, and JBuilder Foundation.  Repast provides various skeletons of agents and their environment, which are open to modification. By extending basic agents' functions, users can build their own new agents to perform complex activities. It also features with parameter management modules for simulation set up. While running the simulation, Repast is able to collect the output and reflect the simulator behavior simultaneously. Afterwards, a collection of data synthesis toolkits are available for simulation result analysis. Repast is discussed as a CRN simulator because of its popularity and some proposed wireless network platforms.

WSN mIddleware Service moDules simulation platform (WISDOM) \cite{Lim:2008aa} is a novel wireless sensor network (WSN) simulation framework based on Repast, which works as the simulation engine to perform discrete event-driven simulations. The simulator primarily consists of four basic components: agents, messengers, configurations, and statistics. Agents are the main components of the network. They can evolve to various identities in the network with different functionalities. Messengers are in charge of information exchange among agents. Configuration module helps for a continent model set-up. Statistic module provides user with result analysis instrument \cite{Lim:2008aa}. The designed middle-ware services for routing, scheduling, agent management, target tracking are valuable on simulating cognitive radios. For example, WISDOM has been used for simulating an enhanced sensor scheduling protocol \cite{Fu:2008aa}.

Recently, Repast is adopted to discuss some marginal CRN studies such as CRN availability in real world \cite{Aloi2016} or spectrum market in CRN deployment \cite{Basaure2015}.

\subsection{NetLogo}
NetLogo is a multi-agent programmable modeling environment running on the Java virtual machine, which can run on different platforms \cite{NetLogo}. As a free, open source software, NetLogo is a multi-purposed simulator that is not exclusively used for computer or communication networks. It is an agent based simulator that is particularly suitable for modeling complex systems developing over time with support of either 2D or 3D visualization. Everything in NetLogo is considered as an agent. There are four different types of agents:

\begin{itemize}
    \item Turtles: moving agent in the world;
    \item Patches: components of the two-dimensional world;
    \item Links: connections between turtles; and
    \item Observer: monitor and recorder of the simulation process.
\end{itemize}

When used for simulation, like other ABM software, NetLogo provides many sample models as demonstration. Users can start to build their model by choosing proper sample model, and modifying it using programming language Logo to meet their requirements. NetLogo offers real time simulation process presentation and result reflection. In current version (5.3.1) of NetLogo, there are many integrated sample models available for modification and use, but no CRN sample model is included. However, some wireless network/radio analysis models, such as Priority\_slots\_wireless\_dynamic, Radio\_wireless, WSN, and more, are handy from community NetLogo model library.

Based on the natural characteristics of NetLogo, researchers made some trials for cognitive radios simulation by considering network components as different agents. In \cite{Sasirekha:2012aa}, a CRN model including one Spectrum Coordinator (SC), several PUs and SUs, was created to test a proactive spectrum sensing scheme. Researchers have discussed experiment details on CRN simulation by using NetLogo in \cite{Dzikowski2015}.  However, as a general purposed simulator for complex dynamic systems, NetLogo faces a lot of limitations when it is used for network simulation due to the lack of network modules and implementations of network protocols.

\subsection{AnyLogic}
As a simulator that supports both discrete event modeling and agent based modeling, AnyLogic has been widely used in complex systems simulation for time scheduling, task distribution, energy management and path optimization. Constructed using Java, AnyLogic can uniformly run on all major platform with devotional supports. As a commercial software, AnyLogic is equipped with many powerful simulation features, such as

\begin{itemize}
    \item large numbers of in-built agent prototypes;
    \item multiple modeling structures, including agent-based, system dynamics, discrete-event, continuous and dynamic system models;
    \item Java support environment;
    \item Rich GUI functionality; and
    \item Good technical supports.
\end{itemize}

To the best of our knowledge, AnyLogic has not been used for CRNs so far. But it has been used for wireless ad-hoc networks modeling and wireless sensor networks (WSNs) modeling \cite{Dressler:2008aa}. However, considering its maturity and the good supports, it is potentially can be considered to simulate cognitive radio networks.

\subsection{Discussions}
In cognitive radio study, agent-based simulators are not as popular as traditional network simulators due to the lack of integrated network structures and protocol stacks. Nevertheless, cognitive radio can be viewed as a combination of a bunch of self-disciplined individuals; each of them is a multi-functional agent when applying to ABM. It is foreseeable that ABM will attract more attention in CRN simulation with development of new network models in these agent-based simulators.

\section{Conclusions}
\label{sec:conclusion}
In this paper, we provide a survey on simulation tools for experimental studies for cognitive radio networks. In general, there is no perfect simulator can meet all-round CRN simulation requirements. Signal processors such as GNU Radio and Matlab is appropriate to study spectrum sensing and analysis on physical layer or in radio environment, while traditional non-agent-based network simulator can be used on large-scale CRN simulation regarding to spectrum management, spectrum sharing, energy management, mobility management, routing protocol, and QoS control. In comparison, agent-based CRN simulation still stays in infant phase, despite some coherent features are shared between cognitive radios and agent-based modeling. Specifically, researchers is encouraged to choose a proper platform or a combination of some for certain purpose.

We hope our analysis and discussions are helpful to researchers who are looking for proper experimental simulation tools for studying CRNs, and also expect the survey can inspire more interests and discussions in the community, and lead to the construction of more advanced CRN simulators.

\section*{ACKNOWLEDGEMENT}
This work is partially supported by the US National Science Foundation, EARS Program under Grant No. CNS-1443885.

\footnotesize

\bibliographystyle{IEEEtranS}
\bibliography{CRN_DQ}

\end{document}